\documentclass[12pt]{amsart}
\usepackage[mathscr]{eucal}
\usepackage{amscd}
\usepackage{amssymb,latexsym}
\newtheorem{thm}{Theorem}%[section]
\newtheorem{lem}[thm]{Lemma}%[section]
\newtheorem{Def}{Definition}%[section]
\newtheorem{prop}[thm]{Proposition}%[section]
\renewcommand\S{\Sigma}
\newcommand\s{\sigma}
\renewcommand\d{\partial}

\renewcommand\L{\triangle}
\newcommand\D{\nabla}
\newcommand\e{\epsilon}
\renewcommand\b{\beta}

\renewcommand\div{\operatorname{div}}

\newcommand\ric{\operatorname{Ric}}

\newcommand\g{\gamma}
\newcommand\8{\infty}
\renewcommand\a{\alpha}
\newcommand\hess{\operatorname{Hess}}

\newcommand\oM{\overline M}
\newcommand\sH{\mathscr{H}}

\newcommand\beq{\begin{eqnarray}}
\newcommand\eeq{\end{eqnarray}}

\newcounter{mnotecount}[section]

\begin{document}
\setlength{\baselineskip}{.51cm}
\title[]
{Boundaries of zero scalar curvature in the AdS/CFT Correspondence}

\author[]{Mingliang Cai and  Gregory J. Galloway}
\thanks{Supported in part by NSF grant \#
    DMS-9803566.}
\address{University of Miami, Department of Mathematics, 
Coral Gables, FL, 33124}
\email{mcai@math.miami.edu \\galloway@math.miami.edu}

%\begin{abstract}
%\end{abstract}

\maketitle

\section{Introduction}

In \cite{WY}, Witten and Yau consider the AdS/CFT correspondence in the context of a Riemannian 
Einstein manifold $M^{n+1}$ of negative Ricci curvature which admits a conformal 
compactification (in the sense of Penrose \cite{P}) with conformal boundary $N^n$. 
As discussed in \cite{WY}, a conformal field theory on $N$ relevant to this correspondence
is stable (with respect to the brane action on $M$)
if the conformal class of the boundary contains a metric of positive scalar curvature 
and is unstable if it contains a metric of negative scalar curvature. 
In the borderline case of zero scalar curvature, the theory may be stable or unstable.
Witten and Yau go on to prove that if the conformal class of $N$ contains 
a metric of positive scalar curvature, then $M$ and $N$
have several desirable properties: (1) $N$ is connected, which avoids the difficulty of 
coupling seemingly independent conformal field theories, (2) the $n$th homology of
$\overline M$ vanishes, in particular $M$ has no wormholes and (3) at the fundamental group level, the
topology of $M$ is ``bounded by" the topology of $N$.  

The aim of the present paper is to show that
all of these results extend to the case where the conformal class of the boundary contains
a metric of nonnegative scalar curvature.  By a well known result of Kazdan and Warner \cite{KW},
if $N$ has a metric of nonnegative scalar curvature, and if the scalar curvature is positive at some
point, then N has a conformally related metric of  positive scalar curvature.  
Hence, the essential case handled here is the case in which the conformal class of the
boundary contains a metric of zero scalar curvature.  
%Thus we may,
%and henceforth shall, restrict attention to the case in which the conformal class of the
%boundary contains a metric of zero scalar curvature. 
The  proof method used in this paper is quite different from, and in some sense
dual to, that used in \cite{WY}.  The method of \cite{WY} involves minimizing the 
co-dimension one brane action on $M$, and uses
the machinery of geometric  measure theory, while the arguments presented here use only geodesic
geometry. We proceed to a precise statement of our results.

Let $M^{n+1}$ be a complete Riemannian manifold, with metric $g$, and suppose $M$ admits a conformal
compactification, with conformal boundary (or conformal infinity) $N$.  Thus it is assumed that
$M$ is the interior of a compact manifold-with-boundary $\oM^{n+1}$ and that there exists a smooth function
$r$ on $\oM$ such that (1) $r>0$ on $M$, (2) $r=0$ and $dr \ne 0$ along $N =\d\oM$, and (3) $r^2g$ extends
smoothly to a Riemannian metric $\bar g$ on $\oM$. The induced metric $h=\bar g|_{TN}$ on $N$
changes by a conformal factor with a change in the \emph{defining function\/} $r$, and so $N$ has a 
well defined conformal structure.  If the conformal class of metrics on $N$ contains a metric 
of positive (resp., nonnegative, zero, etc.) scalar curvature we say that $N$ has positive 
(resp., nonnegative, zero, etc.) scalar curvature.   

In \cite{WY}, Witten and Yau consider conformally compactified orientable
Einstein manifolds $M^{n+1}$ which satisfy
$\ric = -ng$.  More generally, their results allow $\ric \ge -n g$ (i.e., $\ric(X,X) \ge -ng(X,X) = -n$
for all unit vectors $X$) provided $\ric \to -ng$ sufficiently fast as one approaches the conformal boundary
$N$.  
(Regarding this fall-off, it is sufficient to require $\ric = -ng + o(r^2)$, where $r$ is a suitably chosen
defining function for the conformal boundary. This is discussed in more detail below.)  
In this setting they prove that if $N$ has a component of positive scalar curvature then
(1) $N$ is connected, (2) $H_n(\oM,\Bbb Z) =0$, and (3) the map $i_*: \Pi_1(N) \to \Pi_1(\oM)$ 
induced by inclusion $i: N \to \oM$ is onto.   The last result says that any loop in $M$ can be deformed
to a loop in $N$.  Thus, at the fundamental group level, the topology of $M$ can be no more
complicated than the topology of $N$.  In particular, if $N$ is simply connected, so is $M$.  
The following theorem generalizes these results by weakening the scalar curvature condition on $N$.

\begin{thm}\label{th:a} Let $M^{n+1}$ be a complete Riemannian manifold which admits a conformal
compactification, with conformal boundary $N^n$, and suppose the Ricci tensor of $M$ satisfies,
$\ric \ge -ng$, such that $\ric \to -ng$ sufficiently fast on approach to conformal infinity.
If $N$ has a component of nonnegative scalar curvature then the following properties
hold. 
\begin{enumerate}
\item[(a)] $N$ is connected.  
\item[(b)] 	If $M$ is orientable, the $n$th homology of $\oM$ vanishes, $H_n(\oM,\Bbb Z) =0$.
\item[(c)] The map $i_*: \Pi_1(N) \to \Pi_1(\oM)$ ($i$ = inclusion) is onto.
\end{enumerate}
\end{thm}

The essential step in the proof is to establish part (a).  Part (c) then follows from part (a)
by a covering space argument, as noted in \cite{WY}. 
In turn, as will be shown here, part (b) follows from part (c) 
via basic homology theory (essentially Poincar\'e duality; cf. \cite{G1+,G2+}, where similar
arguments have been used).  
In order to give a flavor of the sorts of techniques that will be
used to prove part (a), we will first give a proof of the connectedness of $N$ in the
setting of Witten and Yau \cite{WY}; i.e., under the assumption that 
$N$ has a component of positive scalar curvature.  

The result of Witten and Yau on the connectedness of the boundary is easily derived 
from the following proposition.   

\begin{prop}\label{th:b} Suppose $M^{n+1}$ is a complete Riemannian manifold-with-boundary
having Ricci curvature greater than or equal to $-n$.  If the boundary $\d M$
is compact and has mean curvature $H > n$ then $M$ is compact. 
\end{prop}

By our conventions, $H = \div_{\d M}X$, where $X$ is the outward pointing unit normal along 
$\d M$.  Results  similar to Proposition \ref{th:b} are obtained in \cite{WY} by minimizing
the brane action and making use of the machinery of geometric measure theory.  Here we give a 
proof of Proposition \ref{th:b} using basic techniques in geodesic geometry.  These arguments are
reminiscent of the kinds of arguments used in the proof of the classical Hawking-Penrose singularity
theorems.

\smallskip
\noindent
{\it Proof of Proposition \ref{th:b}:} 
Suppose $M$ is noncompact.  Then we can find a point
$q\in M$ such that the distance from $q$ to $\d M$ is greater than $\coth^{-1}(1+\delta)$.
Here $\delta >0$ is chosen so that the mean curvature $H$ of $\d M$ satisfies $H \ge n(1+\delta)$.
Let $p$ be a point on $\d M$ closest to $q$, and let $\s: [0,\ell]  \to M$ be a unit speed minimal
geodesic from $p=\s(0)$ to $q= \s(\ell)$. 
Let $\rho: M \to \Bbb R$ be the distance function to the boundary,
\beq
\rho(x) = d (x,\d M) = \inf_{y\in \d M} d(x,y)\,  .\nonumber
\eeq  
In general, $\rho$ is  continuous on $M$ and smooth outside the focal cut locus
of $\d M$.  In particular, since $\s$ realizes the distance to $\d M$, $\rho$ is 
smooth on an open set $U$ containing $\s \setminus \{q\}$. 

For $0\le s < \ell$, let $H(s) = -\triangle\rho(\s(s)) = \div(-\nabla\rho)(\s(s))$.
Geometrically, $H(s)$ is the mean curvature of the slice $\rho = s$, with respect to
the unit normal $-\nabla\rho$, at the point $\s(s)$.  $H= H(s)$ obeys the 
well known traced Riccati equation~\cite{Kar},
\beq\label{eq:a}
 H' = \ric(\s',\s') + |B|^2 \, ,
\eeq   
where $'=d/ds$ and $B(s)=- \operatorname{Hess}(\rho)(\s(s)) = \nabla(-\nabla\rho)(\s(s))$.
Since $H(s)$ is the trace of $B(s)$, the Schwarz inequality implies, $|B|^2 \ge H^2/n$.
Equation \ref{eq:a}, taken together with this inequality and the inequalities $\ric(\s',\s')\ge -n$ and
$H(0) = H_{\d M}(p)\ge n(1+\delta)$, implies that $\mathscr H(s):= H(s)/n$ satisfies,
\beq 
\sH'  \ge  \sH^2 - 1, \qquad \sH(0)  \ge  1+\delta.  \nonumber
\eeq
By comparison with the unique solution to: $h' =h^2-1$, $h(0) = 1+ \delta$, we obtain,
\beq
\sH(s) \ge \coth(a-s)\, , \nonumber
\eeq
where $a =\coth^{-1}(1+\delta) < \ell = d(q,\d M)$.  This inequality implies that $\sH = \sH(s)$ is
unbounded on $[0,a)$, which contradicts the fact that it is smooth on $[0,\ell)$. Thus,
$M$ must be compact. 
\endproof

\smallskip

We remark that the rigid version of Proposition \ref{th:b}, in which one assumes the
weak inequality, $H \ge n$, holds, has  previously been treated in the 
literature, cf., \cite{Kas, CK}. 

Return to the setting of Theorem \ref{th:a}, but in the case considered by Witten and Yau 
\cite{WY} in which some component $N_0$, say, of $N$
has positive scalar curvature.  We indicate how 
the connectedness of $N$ in this case follows from Proposition~\ref{th:b}.
%As shown in \cite{GW}, there exists a defining function $r$ for the
%conformal boundary $N$ such that near N, $(M,g)$ takes the form,
%\beq
%M \approx (0,r_0)\times N, \qquad g = \frac1{r^2}(dr^2 + h_r) \, ,  
%\eeq
%where $h_r$ is an $r$-dependent family of metrics on $N$.
Let $r$ be a defining function for the conformal boundary $N$, and
let $U$ be a neighborhood of $N_0$ which does not meet any other components of $N$.
Let $M_t = M\setminus \{x\in U: r(x) < t\}$. For $t$ sufficiently small, $M_t$ is 
a manifold-with-boundary, with boundary $\d M_t$ diffeomorphic to $N_0$, which satisfies      
the hypotheses of Proposition~\ref{th:b}.  (That the mean curvature of $\d M_t$ 
satisfies $H > n$ uses the assumption of positive scalar curvature on $N$. It also uses
the fall-off condition on the Ricci curvature; without that, there are simple counter-examples to
the Witten-Yau result.) 
We conclude that $M_t$ is compact, from which it follows that $N$ has no other
components; i.e., $N=N_0$, and hence is connected.

We now describe  briefly our approach to the proof of part (a) of Theorem~\ref{th:a}. The idea, roughly, is
as follows:    First we show that if there is
a sequence of compact hypersurfaces $\S_k$ in $M$ going
to infinity in some end such that the mean curvature $H_k$ of $\S_k$
approaches $n$ ``fast enough" then $M$ has only one end.
Then we show that this rate 
is  realized if the end admits a conformal
compactification such that the conformal class of the boundary
has a metric of zero scalar curvature.  The first, and main, step is to establish
a suitable refinement of Proposition \ref{th:b}.  

\begin{thm}\label{th:c} Let $M^{n+1}$ be a complete Riemannian manifold 
having Ricci curvature greater than or equal to $-n$.  Fix a base point
$o\in M$.  Suppose there exists a sequence of compact hypersurfaces 
$\{\S_k\}$ satisfying the following conditions.  
\begin{enumerate}
\item[(a)] Each $\S_k$ separates $M$.  We call the component of $M\setminus \S_k$ containing
$o$ the inside of $\S_k$ and the other component the outside.   
\item[(b)] 
%The distance from $o$ to $\S_k$ tend to infinity, 	
$d(o,\S_k) \to \infty$ as $k\to \infty$. 
\item[(c)] Denote by $H_k$ the mean curvature of $\S_k$ with respect to the outward normal,
and let $h_k$ be the smaller of $\min\{H_k(x): x\in \S_k\}$ and n.  Assume that 
\beq\label{eq:b}
\lim_{k\to \8}(n-h_k)e^{2d(o,\S_k)} = 0.
\eeq
\end{enumerate}
Then $M$ has one or two ends.  If $M$ has two ends then $M$ is isometric to $\Bbb R\times \S$,
with warped product metric $dr^2 + e^{2r}g_0$, where $\S$ is compact and $g_0$ is a metric of
nonnegative Ricci curvature on $\S$.
\end{thm}

%The reader unfamiliar with the general definition of an end  
%may take an end to be an unbounded component of 
%the complement of a compact set in $M$ which is diffeomorphic to $\Bbb R\times \S$,
%where $\S$ is a compact hypersurface.  The ends of a conformally compactified manifold
%are of this form.     

An \emph{end} of $M$ is, roughly speaking, an unbounded component of the complement of a sufficiently large
compact subset of $M$, cf. e.g., \cite{A} for a precise definition. If $M$ admits a conformal 
compactification then each of its ends is diffeomorphic to $\Bbb R\times \S$, where $\S$ is a
component of the conformal boundary.   
In the special case of two ends in the theorem,  $M$ has a \emph{cusp} at one end, and
hence does not admit a conformal compactification.  (Regardless of curvature 
conditions, the ends of a conformally
compactified manifold must have positive mean curvature near infinity.)

We recall an example considered in \cite{WY} which satisfies all the hypotheses
of Theorem \ref{th:c} except for the mean curvature decay condition (\ref{eq:b}). 
Let $(\S ,g_0)$ be any compact negatively curved ($\ric = -(n-1) g$) Einstein manifold of dimension
$n$.  Then $M = \Bbb R \times \S$, with warped product metric $g= dr^2+ \cosh^2(r)g_0$   
is an Einstein manifold satisfying, $\ric = -ng$. The slices $\S_r = \{r\} \times \S$, $r>0$, have 
mean curvature $H(r)= n\tanh r$.  Hence $H(r)\to n$ as $r\to\8$, but
$\lim_{r\to\infty}e^{2r}(n - H(r)) =2n$.  This shows that the mean curvature condition in Theorem
\ref{th:c} is in some sense optimal.  
%Similar examples can be constructed satisfying $\ric \ge -ng$
%and having arbitrarily many ends of the same form. 

In the next section we present the proof of Theorem \ref{th:c}, and in the
final section we present the proof of Theorem \ref{th:a}.

\section{Proof of Theorem \ref{th:c}} 

The proof of Theorem \ref{th:c} is similar in spirit
to the proof of the Cheeger-Gromoll splitting theorem \cite{CG, EH}, and makes use 
of (generalized) Busemann functions and the method of support functions \cite{W,EH}.  The method
of support functions provides an elementary way to work with the Laplacian of certain
geometrically defined functions, such as Busemann functions, which  are in general only 
$C^0$.  

Let $M$ be a Riemannian manifold and let $f\in C^0(M)$ be a continuous function on $M$.
A \emph{lower support} function for $f$ at $p\in M$ is a function $\phi$ defined and continuous
on a neighborhood $U$ of $p$ such that $\phi \le f$ on $U$ and $\phi(p) = f(p)$.  We say that
$f\in C^0(M)$ satisfies $\triangle f\ge a$ ($a\in \Bbb R$) in the \emph{support sense} provided 
for each $p\in M$ and every $\e > 0$ there exists a $C^2$ lower support function $\phi_{p,\e}$
for $f$ at $p$ such that $\L \phi_{p,\e}(p) \ge a-\e$.  The Hopf-Calabi maximum principle asserts
that if $M$ is connected and $f \in C^0(M)$ satisfies $\L f\ge 0$ in the support sense then $f$ cannot
attain a maximum unless it is constant.  The proof of the Hopf-Calabi maximum principle is completely
elementary; a short elegant proof is given in  \cite{EH} (cf., also  \cite{B}).  We will need to make use
of a slightly more general version of the Hopf-Calabi maximum principle.

\begin{Def} A function $f\in C^0(M)$ satisfies $\L f\ge a$, $a \in \Bbb R$, in the \emph{generalized
support sense} provided for each $p\in M$, there is a neighborhood $U$ of $p$ such that the following
conditions hold.
\begin{enumerate}
\item[(a)] There exists a sequence $\{f_k\}$, $f_k\in C^0(U)$, such that $f_k \to f$ uniformly
on~$U$.
\item[(b)] $\L f_k \ge a_k$ on $U$ in the  support sense, and $a_k \to a$. 
\end{enumerate}
\end{Def}

\begin{lem}\label{th:d}\emph{(Generalized maximum principle).}
Suppose $M$ is a connected Riemannian manifold, and $f\in C^0(M)$ satisfies 
$\L f \ge 0$ in the generalized support sense.  Then, if $f$ attains a maximum, it is
constant.
\end{lem}

\begin{proof} The proof is a simple modification of the proof of the Hopf-Calabi maximum principle
given in \cite{EH}.  We omit the details.
\end{proof} 

One defines $\L f \le a$ in the generalized support sense in a similar way, using $C^2$
\emph{upper} support functions.  By definition,  $\L f =a$ in the generalized
support sense provided  $\L f \ge a$ and $\L f \le a$ in the generalized support sense.  
If $\L f =a$ in the generalized support sense then $f\in C^{\infty}(M)$ and 
$\L f =a$ in the usual sense.  Indeed, for any small geodesic ball $B$,  basic 
elliptic theory \cite{GT} guarantees that the Dirichlet problem: $\L h =a$, $h|_{\d B} = f|_{\d B}$,
has a  solution $h\in C^{\infty}(B)\cap C^0(\overline B)$.  Then $\L (f-h) = 0$ on $B$ in the
generalized support sense, and the generalized maximum principle applied to $\pm(f-h)$  implies that
$f|_B = h$.

\smallskip
\noindent
\emph{Proof of Theorem \ref{th:c}:}  Suppose that $M$ has more than one end.  Then there is a
compact set $K$ such that $M\setminus K$ has at least two unbounded components $E_1$ and $E_2$,
say.  Since $M\setminus K$ has at most finitely many components we may assume without loss of generality 
that $\S_k \subset E_1$ for all $k$.  We now construct a \emph{line} in $M$.  (Recall, a line 
is a complete unit speed geodesic, each segment of which realizes the distance between its endpoints.)
Let $\{q_k\}$ be a sequence in $E_2$ going to infinity, $d(o,q_k) \to \infty$.  Let $p_k$ be a point 
on $\S_k$ closest to $q_k$, and let $\s_k:[-a_k,b_k]\to M$ be a unit speed minimal geodesic from $p_k$ to
$q_k$. Since  $\s_k$ meets $K$, we may parameterize $\s_k$ so that $\s_k(0)\in K$.  By passing to a
subsequence if necessary we have $\s_k(0) \to \bar o\in M$ and $\s_k'(0) \to X\in T_{\bar o}M$.
Let $\s: \Bbb R \to M$ be the geodesic satisfying $\s(0) = \bar o$ and $\s'(0) = X$.  As $\s$ is the
limit of minimal segments, it is a line in $M$. 

We consider two Busemann functions on $M$, a Busemann function associated with the  sets
$\S_k$ (in the sense decribed in \cite{W}) and the standard Busemann function associated
with the ray (half-line) $\s|_{[0,\8)}$. 
For each $k$, let $\b_k:M \to \Bbb R$ be the function defined by,
$\b_k(x) = d(\bar o, \S_k) - d(x, \S_k)$.  The triangle inequality implies that $\b_k$ is Lipschitz
continuous, with Lipschitz constant one, and satisfies $|\b_k(x)| \le d(x,\bar o)$.  Hence, the family
of functions $\{\b_k\}$ is equicontinuous and uniformly bounded on compact subsets.  Thus, by Ascoli's
theorem, and passing to a subsequence if necessary, we have that $\b_k$ converges on compact subsets
to a continuous function $\b:M\to \Bbb R$, called the Busemann function associated with $\{\S_k\}$.
We will ultimately show that $\b\in C^{\8}(M)$ and satisfies $\L \b = n$, from which 
the special form of $(M,g)$ in the statement of Theorem \ref{th:c}
will readily follow.  The first step is to establish the following.

\smallskip
\noindent
{\bf Claim.} \emph{$\L \b \ge n$ in the generalized support sense}.

\smallskip

Let $p$  be any point in $M$ and let $B=B(p,r)$ be a small geodesic ball centered at $p$
of radius $r$.  To prove the claim we show that $\L \b_k \ge n_k$ on $B$ in the support sense,
where $n_k \to n$.   Given $q\in B$ and $\e>0$, we construct a support function $\b_k^{q,\e}$  
for $\b_k$ at $q$ as follows.  Let $z$ be a point on $\S_k$ closest to  $q$, and let $\g:[0,\ell]:M\to \Bbb
R$ be a unit speed minimal geodesic from $z=\g(0)$ to $q=\g(\ell)$. Let $V$ be a small neighborhood of
$z$ in $\S_k$. By bending $V$ slightly toward the outside of $\S_k$ we obtain a smooth hypersurface
$V'$ with the following properties: (1) $z\in V'$ is the unique closest point in $V'$ to $q$, (2) the
second fundamental form of $V'$ at $z$ (with respect to the outward normal) is strictly less than that of
$V$, and (3) the mean curvature of $V'$ at $z$ satisfies $H_{V'}(z)\ge H_V(z) -\e\ge h_k-\e$. 
By construction, $\g$ minimizes the distance from $q$ to $V'$, and there are no focal cut points
to $V'$ on $\g$ (in particular, $q$ is not a focal cut point).  Hence the function 
$\b_k^{q,\e}(x)= d(\bar o, \S_k) - d(x, V')$ is a  lower support function for $\b_k$ at $q$ which is
smooth on a neighborhood of $\g$.  

For $0\le s\le \ell$, let $H(s) = \L \b_k^{q,\e}(\g(s))$.  
Arguing as in Proposition \ref{th:b}, $\mathscr H(s)= H(s)/n$ satisfies, 
$\sH'  \ge  \sH^2 - 1$, $\sH(0)  \ge  (h_k-\e)/n$.  Since $(h_k-\e)/n < 1$, by comparing 
with the unique solution to  $h' =h^2-1$, $h(0) = (h_k-\e)/n$, we obtain,
$\sH(s) \ge \tanh(a-s)$, where $a = \tanh^{-1}((h_k-\e)/n)= \frac12\ln(\frac{n+h_k-\e}{n-h_k+\e})$.  
Setting $s =\ell$ in this inequality, we obtain,
\beq\label{eq:c}
\L \b_k^{q,\e}(q) = H(\ell) & \ge & n \, \tanh(a-\ell) =n\, \frac{e^{2a}-e^{2\ell}}{e^{2a}+e^{2\ell}}\\
& = & n\,\frac{(n+h_k-\e) -(n-h_k+\e)e^{2\ell}}{(n+h_k-\e) +(n-h_k+\e)e^{2\ell}} \nonumber\, .
\eeq      

Now, by the triangle inequality, $\ell = d(q,\S_k) \le r + d(o,p) + d(o,\S_k)$, 
and hence $e^{2\ell} \le Ce^{2d(o,\S_k)}$, where $C = e^{2(r + d(o,p))}$.  Making use of this latter 
inequality in (\ref{eq:c}) we conclude that $\L \b_k \ge n_k$ on $B$ in the support sense, where
\beq
n_k = n\,\frac{(n+h_k) -C(n-h_k)e^{2d(o,\S_k)}}{(n+h_k) +C(n-h_k)e^{2d(o,\S_k)}}\, .
\eeq 
Invoking the mean curvature condition (\ref{eq:b}), we see that $n_k \to n$.  This yields
the claim.      
  
We now consider the standard Busemann function associated to the ray $\s|_{[0,\8)}$.
For each $s>0$, define the function $b_s:M\to \Bbb R$ by, $b_s(x) = d(\bar o,\s(s)) - d(x,\s(s))
= s - d(x,\s(s))$.  For each $x\in M$, $b_s(x)$ is increasing in $s$ and bounded by $d(\bar o,x)$.
The Busemann function $b:M\to \Bbb R$ of $\s|_{[0,\8)}$ is defined to be the limit function, $b(x) =
\lim_{s\to\8}b_s(x)$.  Because the family $\{b_s\}$ is equicontinuous, $b$ is continuous.  

In the present situation in which the Ricci curvature is greater than or equal to $-n$, it is 
known that $\L b\ge -n$ in the generalized support sense (in fact, in the 
support sense, cf. \cite{CK}). 
As the arguments involved to show this are similar to (but simpler than) the arguments used in the proof
of the claim, we make only a few brief comments.  The relevant support functions $b_s^{q,\e}$
for $b_s$ are defined as follows.  Let $\g:[0,\ell]\to M$ be a unit speed minimal geodesic from
$q$ to $\s(s)$.  The function $b_s^{q,\e}$ defined by, $b_s^{q,\e}(x) = s-(\e+d(x,\g(\ell-\e))$ 
is a lower support function for $b_s$ at $q$ which is smooth near $q$.  By standard comparison
techniques \cite{Kar} like those used above, $b_s^{q,\e}$ satisfies $\L b_s^{q,\e}(q)\ge -
n\coth(\ell-\e)$.  
From this it easily follows that $\L b\ge -n$ in the generalized support sense.  

To summarize, we have shown that the Busemann functions $\b$ and $b$ satisfy 
$\L \b\ge n$ and $\L b\ge -n$ in the generalized support sense.  Hence the
sum $f=\b+b$ satisfies $\L f\ge 0$ in the generalized support sense. Moreover,
$f$ satisfies,  $f\le 0$ on~$M$.  Indeed, we have,
\beq\label{eq:d}
\b(x) + b_s(x) & = & \lim_{k\to\8} [d(\s(0),\S_k) - d(x,\S_k)] + s - d(x,\s(s))\\
& = & \lim_{k\to\8} [d(\s_k(0),\S_k) - d(x,\S_k)] + \lim_{k\to\8}[s - d(x,\s_k(s))]  \nonumber\\
& = & \lim_{k\to\8}[ d(\s_k(0),\S_k) + s - (d(\s_k(s), x)+ d(x,\S_k))] \, . \nonumber
\eeq 

By the triangle inequality, 
\beq\label{eq:e}
d(\s_k(s), x)+ d(x,\S_k) \ge d(\s_k(s), \S_k) = d(\s_k(0),\S_k) + s\, .
\eeq
The inequalities (\ref{eq:d}) and (\ref{eq:e}) imply $\b(x) + b_s(x)\le 0$.  Letting 
$s\to \8$, we obtain $f=\b + b\le 0$.  But note that $f(\bar 0) = \b(\s(0))+ b(\s(0)) = 0+0 =0$.
Thus, by the generalized maximum principle $f\equiv 0$.  Hence, $\b = -b$ and so satisfies
$\L \b \le n$ in the generalized support sense.  Since $\L \b \ge n$ in the generalized support sense,
as well, we conclude from the discussion after Lemma \ref{th:d} that $\b$ is smooth and satisfies
$\L \b =n$ in the usual sense.   

It is known that Busemann functions, where differentiable, have unit gradient, and
hence $|\D\b| = 1$ everywhere. (Briefly, this follows from the fact that $\b$ satisfies, $|\b(q)-\b(p)|\le
d(p,q)$, with equality holding when $p$ and $q$ are on an \emph{asymptotic ray}, cf. Lemma 6 in \cite{W}.) 
This has as well known consequences the fact that $\D\b$ is a geodesic vector field, i.e. its integral
curves are unit speed geodesics, and that $\b$ satisfies (\cite{CG,EH}),
\beq\label{eq:f}
\D\b(\L \b)= \ric(\D\b,\D\b) + |\hess \b|^2\, .
\eeq
(Compare with Equation \ref{eq:a}.)
Since the left hand side vanishes, we have $|\hess \b|^2 = -\ric(\D\b,\D\b)\le n$.
But from the Schwarz inequality, $|\hess \b|^2\ge |\L \b|^2/n =n$.  Hence,
equality holds, which implies,
\beq\label{eq:g}
\hess \b|_{\D\b^{\perp}} = g|_{\D\b^{\perp}}\quad \mbox{ and }\quad \ric(\D\b,\D\b) = -n \, .
\eeq

Exponentiating out from the slice $\S= \b^{-1}(0)$ along its normal geodesics ($=$
integral curves of $\b$) establishes a global diffeomorphism $M\approx \Bbb R \times \S$,
with respect to which $g$ takes the form
\beq
g = dr^2 + g_{ij}(r,x)dx^idx^j \, ,
\eeq
where $\d/\d r= \D\b$ and $g_r = g_{ij}(r,x)dx^idx^j$ is the induced metric on the slice $\S^r =
\b^{-1}(r)\approx \{r\}\times \S$.  Along $\S^r$, $\hess \b(\d_i,\d_j) = \frac12\d_rg_{ij} =g_{ij}$ (by the
first equation in  (\ref{eq:g})), which gives, 
\beq
g = dr^2 + e^{2r}g_{ij}(0,x)dx^idx^j \, ,
\eeq  
as required.  The second equation in (\ref{eq:g}) and a calculation show that $g_0 =
g_{ij}(0,x)dx^idx^j$     is a metric of nonnegative Ricci curvature.  Finally, $\S$ is compact, otherwise
$M\approx \Bbb R\times\S$ has only one end, contrary to assumption.  This concludes the proof of Theorem
\ref{th:c}.
\endproof 

\section{Proof of Theorem 1}  Let $N_0$ be the component in the statement of the theorem which admits 
in its conformal class a
metric $h$ of nonnegative scalar curvature.  As discussed in the introduction, we may assume, in fact, that
$h$ is a metric of zero scalar curvature.  Then there is a defining function 
$r$ such that near $N_0$, $M$ has the form,
$M = [0,r_0)\times N_0$, with metric $g$ of the form,
\beq
g=\frac1{r^2}\bar g = \frac1{r^2} (\frac1{\bar g(\bar\D r,\bar\D r)} dr^2 + g_r) \,\nonumber ,
\eeq
where $g_r$ is the metric induced on $N_r = \{r\} \times N_0$ from $\bar g$, such that $g_0 = h$.
Assume that $(M,g)$ is Einstein, with $\ric = -ng$, or more generally that $(M,g)$ satisfies,
$\ric \ge -ng$ such that, as part of our fall-off assumption, the scalar curvature $S$ of $(M,g)$
satisfies,
\beq
S\to -n(n+1) \mbox{ as } r\to 0 \,\nonumber .
\eeq
As a computation shows, this implies that $\bar g(\bar\D r,\bar\D r)=1$ along $N_0$.  Then
as described in \cite{GW} (cf., Lemma 2.1) the defining function $r$ can be chosen uniquely
in a neighborhood of $N_0$ so that $\bar g(\bar\D r,\bar\D r)=1$ in this neighborhood.
Thus, we may assume that in $[0,r_0)\times N_0$, $g$ has the form,
\beq
g= \frac1{r^2} (dr^2 + g_r) \, . 
\eeq 

With respect to this distinguished defining function we impose the following fall-off
requirement,
\beq\label{eq:h}
r^{-2}(Ric+ng) \to 0 \mbox{ uniformly as } r\to 0.
\eeq
This condition is compatible with the fall-off condition considered in \cite{AM}. Let us also
emphasize that these fall-off conditions are automatically satisfied when $M$ is Einstein 
with $\ric = -ng$. 

The Gauss equation in $(M,g)$ implies,
\beq\label{eq:i}
H^2 = \hat S - S +\ric(X,X) + |B|^2 \, ,
\eeq
where, for each $r$, $\hat S$, $B$ and $H$ are, respectively, the scalar curvature, second fundamental
form and mean curvature of $N_r$, and $X=-r\d/\d r$  is the outward unit normal to $N_r$.
For each $r$, let $\bar S$ denote the scalar curvature of $N_r$ in the metric $g_r$; $\bar S$ and
$\hat S$ are related by $\hat S =r^2 \bar S$. Using this and  the inequality $|B|^2\ge H^2/n$, 
(\ref{eq:i}) implies the  inequality,
\beq\label{eq:j}
n^2-H^2 \le -\frac{n}{n-1}(r^2\bar S + \kappa)\, ,
\eeq
where $\kappa = 2\ric(X,X)-S - n(n-1)$.  
It follows from (\ref{eq:h}) that $r^{-2}\kappa \to 0$ uniformly as $r\to 0$.   
Since $H>0$ for $r$ sufficiently small, we have
$n^2-H^2\ge n(n-H)$ at points where $H\le n$. This, together with (\ref{eq:j}) implies,
\beq\label{eq:k}
r^{-2}(n-H)\le  -\frac{1}{n-1}(\bar S + r^{-2}\kappa) \, \quad\mbox{ where } H\le n \, . 
\eeq 

Pick a sequence $r_k\to 0$, and set $\hat N_k = N_{r_k}$.  
Given $o\in \hat N_{k_0}$, we have,\
\beq
e^{d(o,\hat N_k)} = e^{\int_{r_{k}}^{r_{k_0}} \frac1{r}dr} = r_{k_0}r_k^{-1} \, .
\eeq 
Let $(r_k,x_k)\in \hat N_k$ be a point where the mean curvature of $\hat N_k$ achieves a minimum. 
We may assume 
this minimum mean curvature $h_k = H(r_k,x_k)$ is less than or equal to $n$, otherwise
by the Witten-Yau result we are done.
By passing to a subsequence we may further assume $(r_k,x_k) \to (0,x_0)\in N_0$.    Then, setting $(r,x) =
(r_k,x_k)$ in (\ref{eq:k}) we obtain,
\beq
e^{2d(o,\hat N_k)}(n-h_k) \le -\frac{r_{k_0}^2}{n-1}(\bar S(r_k,x_k) +r_k^{-2}\kappa(r_k,x_k)) \, .
\eeq
Since, as $k\to \8$, $\bar S(r_k,x_k) \to \bar S(0,x_0) = 0$ and $r_k^{-2}\kappa(r_k,x_k)\to 0$,
we have that condition (\ref{eq:b}) in Theorem \ref{th:c} is satisfied.
Then by Theorem \ref{th:c} and remarks in the paragraph following its statement, $M$ has only one
end and hence $N=N_0$, i.e., $N$ is connected.  

This concludes the proof of part (a).  
Part (c)  follows from part (a), just as in the proof of Theorem 3.3 in \cite{WY}. 
%(cf., also, the proof of Theorem 2.2 in \cite{G1+}) 
One passes to the covering space $\overline M'$ of 
$\overline M$ associated with the subgroup $i_*(\Pi_1(N))$ of $\Pi_1(\overline M)$.
The boundary
$\d\overline M'$ contains a copy of $N$ and has more than one component if
$i_*: \Pi_1(N)\to \Pi_1(\overline M)$ is not onto, contradicting 
part (a) applied to $\overline M'$.      
Part~(b)  follows from part (c)
by some basic homology theory, as we now describe.  Similar arguments have been 
used in \cite{G1+,G2+} where related results in the spacetime setting have been 
obtained.  
%(Roughly speaking, it is shown there, as a consequence of
%topological censorship,  that 
%a component of the region of space outside any blackholes in an asymptotically locally
%anti-de Sitter spacetime has only one end.  The result requires a natural energy condition,
%but does not require any assumptions about the curvature of conformal infinity.)

%A related result in the spacetime setting has been
%presented in \cite{G1,G2}.  Roughly speaking, it is shown there, as a consequence of
%topological censorship,  that 
%a component of the region of space outside any blackholes in an asymptotically locally
%anti-de Sitter spacetime has only one end.  The result requires a natural energy condition,
%but does not require any assumptions about the curvature of conformal infinity.  Results
%similar to parts (b) and (c) of Theorem \ref{th:a} are also obtained in \cite{G1+,G2+}.  The sort of
%arguments presented there are applicable here, as well.     

To prove part (c)  consider the relative homology sequence for
the pair $\overline M\supset N$ (all homology is over $\Bbb Z$),
\beq\label{eq:m}
\cdots\quad\rightarrow H_1(N){\buildrel\alpha\over\rightarrow}
H_1(\overline M){\buildrel\beta\over\rightarrow} H_1(\overline M ,N)
{\buildrel\partial\over\rightarrow}{\tilde H_0}(N)=0\quad  \, .
\eeq
(Here $\tilde H_0(N)$ is the reduced zeroth dimensional homology
group.)   To make use of part (c), we use the fact that the first integral 
homology of a space is isomorphic to the fundamental group modded out by
its commutator subgroup.  Hence, modding out by the commutator subgroups
of $\Pi_1(N)$ and $\Pi_1(\overline M)$, we obtain 
a surjective linear map from $H_1(N)$ to $H_1(\overline M)$, i.e.,
$\a$ in (\ref{eq:m}) is onto.  Since $\a$ is onto,
$\ker \beta ={\rm im}\alpha=H_1(\overline M)$ which implies $\beta
\equiv 0$. Hence $\ker\partial = {\rm im}\beta =0$, and thus $\partial$
is injective. This implies that $ H_1(\overline M,N)=0$.
But by Poincar\'e duality for manifolds-with-boundary,
$H_1(\overline M, N) \cong H^{n}(\overline M)\cong H_{n}(\overline M)$,
where for the second isomorphism we have used the fact that $H_n(\overline M)$
is free (cf., \cite{M}).  We conclude that $H_n(\overline M,\Bbb Z) = 0$.

If $M$ is nonorientable (which, by part (a) and a covering space argument,
can happen if and only if $N$ is nonorientable), essentially the
same argument shows $H_n(\overline M, \Bbb Z/2)$ vanishes.

\providecommand{\bysame}{\leavevmode\hbox to3em{\hrulefill}\thinspace}

\end{document}